\documentclass[12pt,a4paper]{article}
\usepackage[utf8]{inputenc}
\usepackage{amsmath}
\usepackage{amsfonts}
\usepackage{amssymb}
\usepackage{makeidx}
\usepackage{graphicx}
\usepackage{float}
\usepackage{hyperref}
\usepackage[toc,page]{appendix}
\usepackage{listings}

\graphicspath{{./}}

\usepackage{amsthm}
\usepackage{mathrsfs}

\newtheorem{prop}{Proposition}[section]

\theoremstyle{remark}
\newtheorem{remark}{Remark}[section]
\theoremstyle{definition}
\newtheorem{defn}{Definition}[section]

\newcommand{\R}{\mathbb{R}}

\allowdisplaybreaks

\begin{document}

\title{\textbf{Critical link of self-similarity and visualisation for jump-diffusions driven by $\alpha$-stable noise}}
\author{Jiao Song and Jiang-Lun Wu \\
{\small\it Department of Mathematics, Swansea University, Swansea SA2 8PP, UK} \\
{\small Email: jiao.song@swansea.ac.uk; j.l.wu@swansea.ac.uk}  }

\date{}

\maketitle

\begin{abstract}
The purpose of this paper is to derive a critical link of parameters for the self-similar trajectories of jump-diffusions which are described as solutions of stochastic differential equations driven by $\alpha$-stable noise. This is done by a multivariate Lagrange interpolation approach. To this end, we utilise computer simulation algorithm in MATLAB to visualise the trajectories of the jump-diffusions for various combinations 
of parameters arising in the stochastic differential equations.
 \end{abstract}



\section{Introduction}
With the passage of time, modelling time evolution uncertainty by stochastic differential equations (SDEs) appears in many diverse areas such as studies of dynamical particle systems in physics, 
biological and medical studies, engineering and industrial studies, as well as most recently micro analytic studies in mathematical finance and social sciences. Beyond modelling uncertainty by Gaussian 
or normal distributions, there is a large amount of data featured with heavy-tailed distributions. On the other side, it is necessary to admit symmetry for the mean (average) by using Gaussian models while asymmetry and/or skewness are accepted by non-Gaussian models. In some applications, asymmetric or heavy-tailed models are needed or even inevitable, in which  a model using stable distributions 
could be a viable candidate. Another important feature of such non-Gaussian models is the use of probability distributions with infinite moments which turns to be more realistic than Gaussian models from 
the view point of heavy tail type data (cf. e.g. \cite{samoradnitsky1994stable}). The research on modelling uncertainty using stable distributions and stable stochastic processes have been increased dramatically, see e.g. \cite{giacometti2007stable}, \cite{zopounidis2013managing}, \cite{dror2002modeling} and \cite{fiche2013features}. The self-similarity property of stable distributions has drawn more and more attention from both theoretical and 
practical view points, i.e \cite{campbell1997econometrics,mandelbrot} and \cite{Zolotarev,leland1993self,shlesinger}.  We refer the reader to \cite{DuWuYang} for discussions of utilising $\alpha$-stable 
distributions to model the mechanism of Collateralised Debt Obligations (CDOs) in mathematical finance.   

Historically, probability distributions with infinite moments are also encountered in the study of critical phenomena. For instance, at the critical point one finds clusters of all sizes while the mean of the 
distribution of clusters sizes diverges. Thus, analysis from the earlier intuition about moments had to be shifted to newer notions involving calculations of exponents, like e.g. Lyapunov, spectral, fractal 
etc., and topics such as strange kinetics and strange attractors have to be investigated. It was Paul L\'evy who first grappled in-depth with probability distributions with infinite moments. Such distributions 
are now called L\'evy distributions. Today, L\'evy distributions have been expanded into diverse areas including turbulent diffusion, polymer transport and Hamiltonian chaos, just to mention a few. Although 
L\'evy's ideas and algebra of random variables with infinite moments appeared in the 1920s and the 1930s (cf. \cite{levy1,levy2}), it is only from the 1990s that the greatness of L\'evy's theory became much 
more appreciated as a foundation for probabilistic aspects of chaotic dynamics with high entropy in statistical analysis in mathematical modelling (cf. \cite{samoradnitsky1994stable,shlesinger}, see also  \cite{mandelbrot,Zolotarev}). Indeed, in statistical analysis, systems with highly complexity and (nonlinear) chaotic dynamics became a vast area for the application of L\'evy processes and the phenomenon 
of dynamical chaos became a real laboratory for developing generalisations of L\'evy processes to create new tools to study nonlinear dynamics and kinetics. Following up this point, SDEs driven by L\'evy 
type processes, in particular $\alpha$-stable noise, and their influence on long time statistical asymptotic will be unavoidably encountered. 

The study of SDEs driven by L\'evy processes is well presented in the monograph \cite{applebaum2009lvy}. Numerical solutions and simulations of $\alpha$-stable stochastic processes were carried out 
in \cite{janicki}. The motivation of this paper is to obtain a critical link among the parameters in the SDEs driven by $\alpha$-stable noises towards self-similarity property from simulations. This can be further linked to sample data analysis after model identifications. We mainly focus on testing two simple types of SDEs, one class is the SDEs with linear drift coefficient and additive $\alpha$-stable noise and the solutions are  called $\alpha$-stable Ornstein-Uhlenbeck processes and the other class is  the linear SDEs (i.e., SDEs with linear drift and diffusion coefficients or the linear SDEs with multiplicative $\alpha$-stable noise) and the solutions are called $\alpha$-stable geometric L\'evy motion. 

\section {Preliminaries}
Given a probability space $(\Omega,\mathcal{F},P)$ endowed with a complete filtration $\{\mathcal{F}_t\}_{t\ge0}$. We are concerned with the following stochastic differential equation (SDE) driven by 
$\alpha$-stable L\'evy motion 

$$dX_t=b(X_t)dt+\sigma(X_t)dB_t+c(X_{t-})dL_t$$  
where $b,\sigma,c:\mathbb{R}\to\mathbb{R}$ are measurable coefficients, $\{B_t\}_{t\ge0}$ is an $\{\mathcal{F}_t\}$-Brownian motion, and $\{L_t\}_{t\ge0}$ is an $\alpha$-stable 
$\{\mathcal{F}_t\}$-L\'evy process with the following L\'evy-Ito representation 

$$L_t=\int_o^{t+}\int_{\vert z\vert<1}\gamma (s-,z)\tilde N(ds,dz)+\int_0^{t+}\int_{\vert z\vert\geq 1}\gamma(s-,z)N(ds,dz)$$ 
with $N:\mathcal{B}([0,\infty)\times\mathbb{R}\setminus\{0\})\to\mathbb{N}\cup\{0\}$ being the Poisson random (counting) measure on $(\Omega,\mathcal{F},P)$ and 
 
$$\tilde N(dt,dz):=N(dt,dz)-\frac{dtdz}{\vert z\vert^{1+\alpha}}$$ 
the associated compensated martingale measure with density $\mathbb{E}N(dtdz)=\frac{dtdz}{\vert z\vert^{1+\alpha}}$, where $\alpha\in(0,2)$ is fixed and 
$\gamma:[0,\infty)\times\mathbb{R}\setminus\{0\}\times\Omega\to\mathbb{R}\setminus\{0\}$ is a c\'adl\'ag (i.e., right continuous with left limits) stochastic process.  

Under the usual conditions, like linear growth and local Lipschitz conditions, for the coefficients  $b,\sigma,c$, there is a unique solution to the above SDE with initial data $X_0$ (see, e.g., \cite{applebaum2009lvy}). In what follows, we introduce two simple ctypes of SDEs fulfilling the usual conditions.

\subsection{The $\alpha$-stable Ornstein-Uhlenbeck processes}

The $\alpha$-stable Ornstein-Uhlenbeck processes are solutions of the following type SDEs 

\begin{equation}\label{OUSDE}
dX_t=-\lambda X_t dt+dL_t
\end{equation}
for $ \lambda>0$, where the $\alpha$-stable noise $dL_t$ is formulated as follows 
$$dL_t=\int_{\vert z\vert<1}\gamma (t-,z)\tilde N(dt,dz)+\int_{\vert z\vert\geq 1}\gamma(t-,z)N(dt,dz).$$
By It\^o formula (cf., e.g., \cite{applebaum2009lvy}), the solution is explicitly given as follows 
\begin{equation}\label{solveOU}
\begin{split}
X_t=&e^{-\lambda t}X_0+e^{-\lambda t}\int^{t+}_0\int_{\vert z\vert<1}e^{\lambda t}\gamma(s-,z)\tilde N (ds,dz)\\
&+e^{-\lambda t}\int^{t+}_0\int_{\vert z\vert\geq 1}e^{\lambda t}\gamma(s-,z)N(ds,dz). 
\end{split}
\end{equation}

\subsection{The $\alpha$-stable geometric L\'evy motion}
Consider the following linear SDE 
$$dX_t=\alpha X_{t}dt+\beta X_{t}dB_t+X_{t-}dL_t$$
where $\alpha>0$, $\beta>0$. 
Then by It\^o formula, one can derive the following explicit solution 
\begin{equation}\label{solveGBM}
\begin{split}
X_t=&X_0\exp\{(\alpha-\frac{1}{2}\beta^2)t+\beta B_t+\int^{t+}_0\int_{\vert z\vert\geq 1}\text{ln}|1+\gamma(s-,z)|N(ds,dz)\\
&+\int^{t+}_0\int_{\vert z\vert<1}\text{ln}|1+\gamma(s-,z)|\tilde N(ds,dz)\\ 
&+\int^t_0\int_{\vert z\vert<1}[\text{ln}|1+\gamma(s,z)|-\gamma(s,z)]\frac{dz}{|z|^{1+\alpha}}ds\}. 
\end{split}
\end{equation}
Due to the above expression, the solution is called an $\alpha$-stable geometric L\'evy motion. 

\subsection{Trajectories and self-similarity}
By applying simulation methods in MATLAB, sample trajectories can be generated and codes are listed in Appendix.\ref{code}. Following graphs show sample trajectories $\alpha$-stable OU processes and $\alpha$-stable geometric L$\acute{e}vy$ motions respectively with a number of parameters combinations.
\begin{figure}[H]
\centering
\includegraphics[width=1\textwidth]{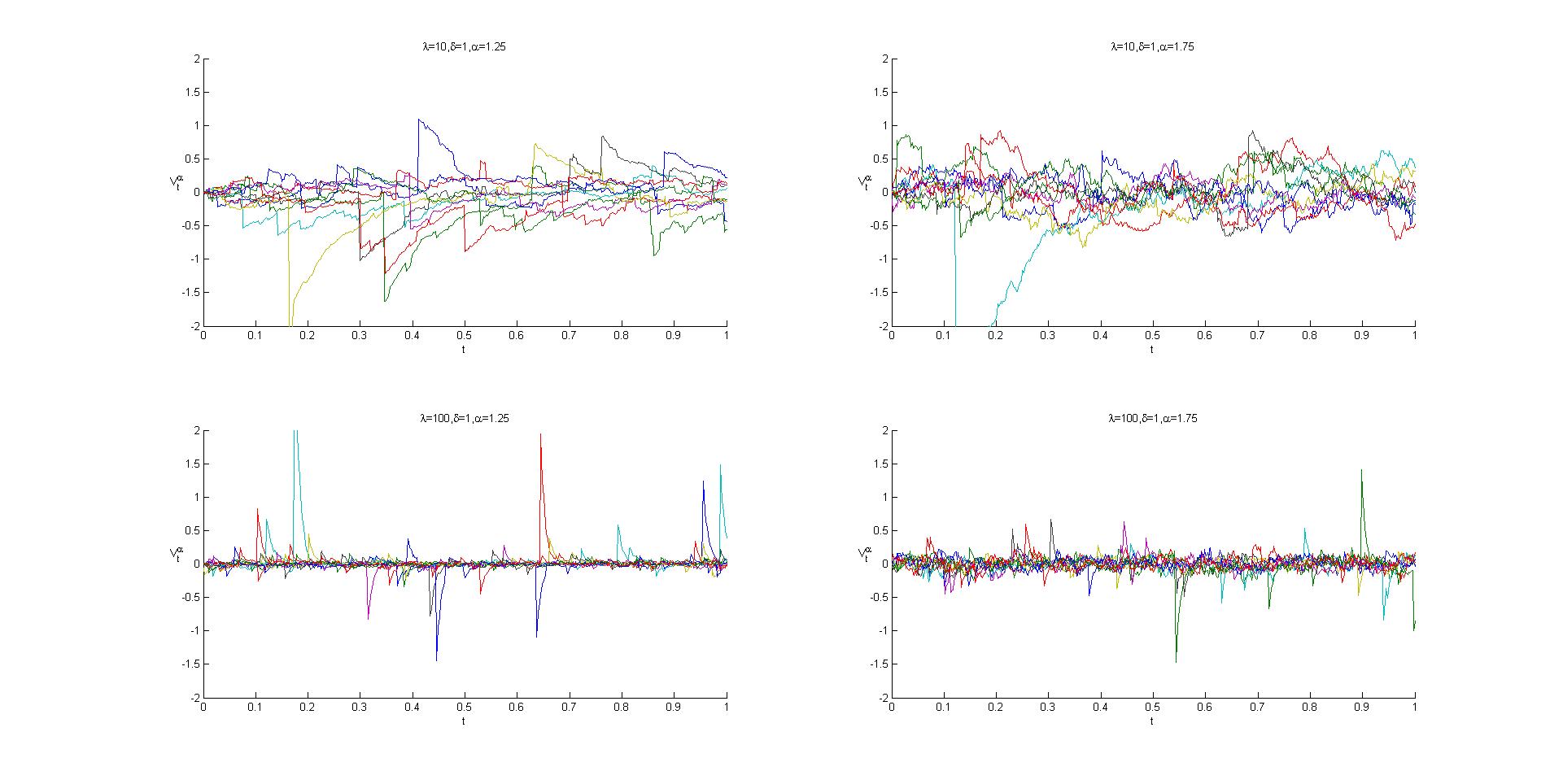}
\caption{$\alpha$-stable OU processes}
\end{figure}
\begin{figure}[H]
\centering
\includegraphics[width=1\textwidth]{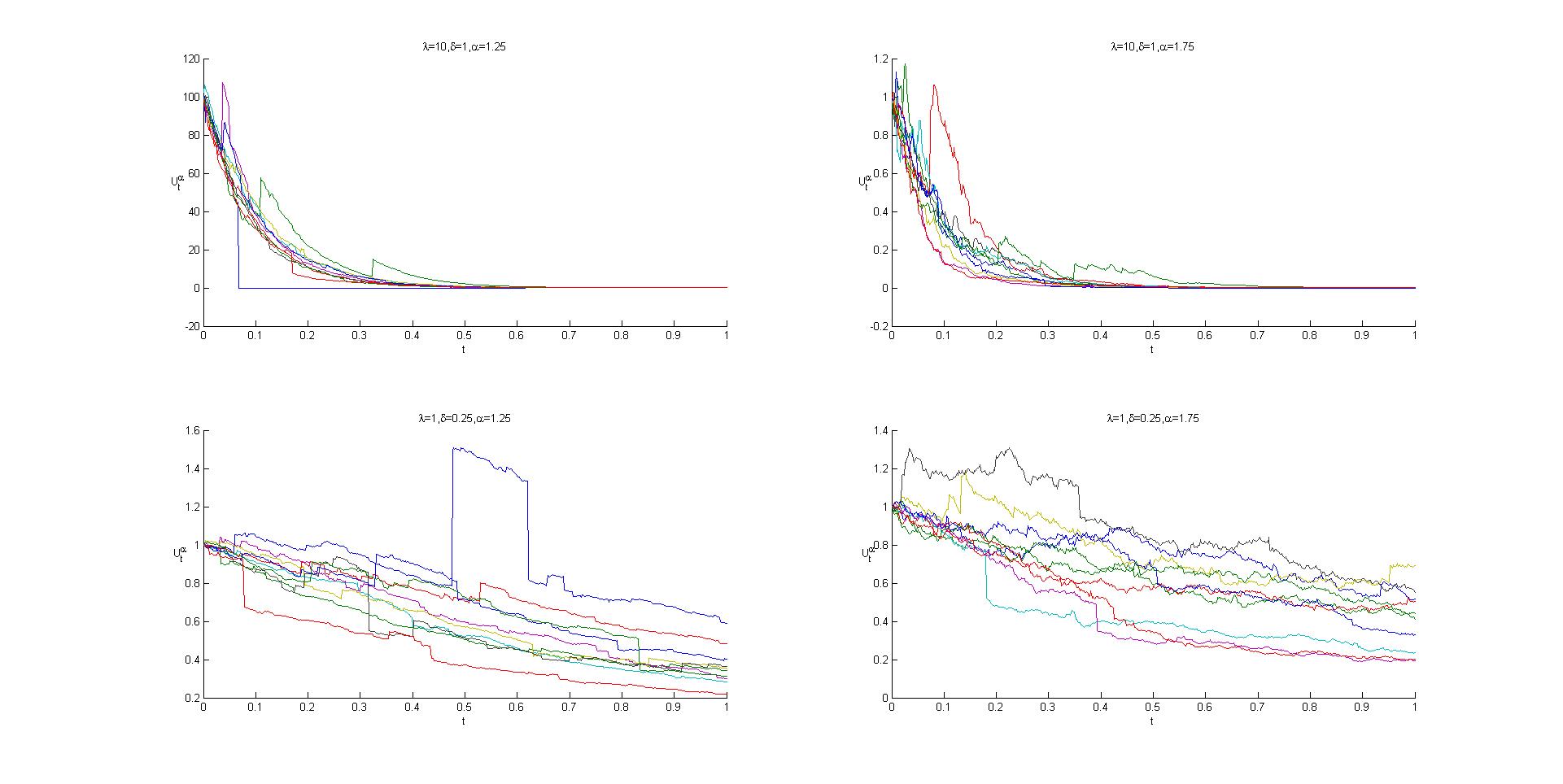}
\caption{$\alpha$-stable geometric L$\acute{e}vy$ motions}
\end{figure}
From the trajectories, self-similarity can be observed. To define self-similarity,
\begin{defn}
 A stochastic process $\{X_t\}_{t\geq 0}$ is said to be ''self-similar'' if for any $a>0$, there exists $b>0$ such that
$$X_{at}\buildrel d \over =bX_t.$$
\end{defn}
For an $\alpha$-stable L$\acute e$vy motion, if we have a real number $c>0$, then the processes $\{X_{ct}\}_{t\geq 0}$ and $\{c^{1/\alpha} X_t\}_{t\geq 0}$ have the same finite dimensional distributions \cite{samoradnitsky1994stable}. Our aim is to obtain a critical link among parameters in SDE towards similarity and it can be used for data fitting purpose in future.

\section{Interpolation}
According to our research problem, polynomial interpolation approach is needed for determining the links among parameters and coefficients in the $\alpha$-stable driven SDEs. The methodology we use is the n-state polynomial interpolation with multiple variables, \cite{de1990multivariate}, \cite{saniee2008simple}, \cite{calvi2005lectures},\cite{liang2006multivariate} and \cite{sauer1995case}.
\\
Let $f=f(x_1,\cdots,x_m)$ be an $m$-variable multinomial function of degree $n$. In \cite{saniee2008simple}, let $\rho=(^{n+m}_n)$ be the number of terms in $f$, and assume we have at least $\rho$ distinct points $\{(x_{1,i},f(x_{1,i})), (x_{2,i},f(x_{2,i}))\cdots(x_{m,i},f(x_{m,i}))\}\in\R^{m+1}$, $1\leq i\leq\rho$, for $f$ to be uniquely defined,
$$f(X_1,\cdots,X_m)=\sum_{e_i\cdot 1\leq n}\alpha_{e_i}X^{e_i}$$
where $\alpha_{e_i}$ are the coefficients in $f$, and also $X=(X_1,\cdots,X_m)$ is the $m$-tuple of independent variables of $f$. $e_i=(e_{1,i},\cdots,e_{m,i})$ is an exponent vector with nonnegative integer entries which has an ordered partition of an integer in [0, n]. $e_i\cdot 1=\sum^m_{j-1}e_{ji}$ stands for vector dot product and $X^{e_i}=\prod^m_{j=1}X^{e_{ji}}_j$. Similarly, comparing with Lagrange interpolation, $\sum^\rho_{i=1}f_il_i(X)$ is the ideal form we would like $f$ to have and $l_i(X)$ is a multinomial function with $X_1,\cdots,X_n$ independent, where X represents the value of $i^{th}$ data. Let us think about a linear equation system
$$f_i=\sum_{e_j\cdot 1\leq n}\alpha_{e_j}X^{e_j}_i$$
where $1\leq i\leq\rho$, now consider,
\begin{equation}
M=[X^{e_j}_i]=
\begin{bmatrix}
   X_1^{e_1} & \cdots & X_1^{e_\rho} \\
   \vdots& &\vdots\\
    X_i^{e_1} & \cdots & X_i^{e_\rho}\\
   \vdots& &\vdots\\
     X_\rho^{e_1} & \cdots & X_\rho^{e_\rho}
  \end{bmatrix}
\end{equation}
as a sample matrix and assume $\det(M)\ne 0$.
\begin{remark}
We want to determine $f$ without solving for its coefficients individually.
\end{remark}
The algorithm is to make some substitutions. If we have $\bigtriangleup=det(M)$, we use $x_j=X$ in M, then we have
\begin{equation}
M_j(X)=
\begin{bmatrix}
   x_1^{e_1} & \cdots & x_1^{e_\rho} \\
   \vdots& &\vdots\\
    X^{e_1} & \cdots & X^{e_\rho}\\
   \vdots& &\vdots\\
     x_\rho^{e_1} & \cdots & x_\rho^{e_\rho}
  \end{bmatrix}\gets j^{th} \text{row}.
\end{equation}
Use $\bigtriangleup_j(X)=\det(M_j(X))$ and $X=x_i$ in $M_j(x)$ where $i\ne j$, then we have
\begin{equation}
(M_j)_i=
\begin{bmatrix}
   x_1^{e_1} & \cdots & x_1^{e_\rho} \\
   \vdots& &\vdots\\
    x_i^{e_1} & \cdots & x_i^{e_\rho}\\
    \vdots& &\vdots\\
    x_i^{e_1} & \cdots & x_i^{e_\rho}\\
   \vdots& &\vdots\\
     x_\rho^{e_1} & \cdots & x_\rho^{e_\rho}
  \end{bmatrix}.
\end{equation}
We can easily see that the $i^{th}$ row appears twice in $(M_j)_i$ which results in $\det((M_j)_i)=0$. Now $X=x_i\Rightarrow \bigtriangleup_i(X)=\bigtriangleup.$ So
\begin{equation}
l_i(X)=\frac{\bigtriangleup_i(X)}{\bigtriangleup}
\end{equation}
and
\begin{equation}
f=\sum^\rho_{i=1}f_i\frac{\bigtriangleup_i(X)}{\bigtriangleup}.
\end{equation}

\section{Simulations and examples}
In this section, a critical link will be obtained for $\alpha$-stable Ornstein-Uhlenbeck process and $\alpha$-stable geometric L$\acute e$vy motion respectively towards self-similarity from simulations by interpolation method introduced in the above section.
Trajectories of $\alpha$-stable Ornstein-Uhlenbeck process as Equation.\ref{OUSDE} with different combinations of parameters in its SDE are included in Appendix.\ref{trajec1}. And the case for $\alpha$-stable geometric L$\acute e$vy motion can be found in  Appendix.\ref{trajec2}. We could clarify the model into different perspectives by observations and general characteristics of trajectories are summarized as follows,
\begin{enumerate}
\item Fix $\lambda$ and $\mu$, the trajectories $\{X_t\}_{t\geq 0}$ become more tempered as the stability index $\alpha$ increases, but the jump size becomes smaller and smaller so that the trajectories become less and less volatile. In other words, for smaller stability index $\alpha$, the trajectories of $\{X_t\}_{t\geq 0}$ are generally more tough than those of bigger stablility index $\alpha$.
\item  Fix $\mu$ and $\alpha$, trajectories look more likely deterministic exponential paths along with the increase of $\lambda$. As for bigger $\alpha$, the trajectories are chaotic more sharply.
\item Fix $\lambda$ and $\alpha$, increasing the volatility parameter $\mu$ indicates higher chaoticity.
\end{enumerate}
\subsection{$\alpha$-stable Ornstein-Uhlenbeck process}
For the triple $(\lambda, \mu,\alpha)$, there is a critical link of the three parameters $\lambda$, $\mu$ and $\alpha$ towards the similarity of trajectories. By simulations, we choose the situations for shows similarity property and keep records of parameters $\lambda$, $\mu$ and $\alpha$ when the first jump appears. Especially, the degree 1 linear relationship among these three parameters is useful in data modelling for uncertainty related problems in reality.
\begin{center}
 \begin{tabular}{||c  c  c  c  c||}
 \hline
 $\lambda$&$\mu$& $\alpha$& t& $X^\alpha_t$ \\ [0.5ex]
 \hline
 1&0.25&1&0.06055&0.4198\\
 \hline
 1&1&1.75&0.003906&-0.1551\\
 \hline
1&100&0.75&0.03125&18.82\\
 \hline
10&0.25&0.5&0.02148&0.4561 \\
 \hline
1000&0.25&1.75&0.001952&0.0374\\[1ex]
 \hline
\end{tabular}
\end{center}
We have degrees $n=1$, variables $m=4$, so terms= $\Big(
\begin{array}{cc}
1+4 \\
1\\
\end{array}\Big)=5.$ If we have $g=f(a,b,c,d)$ which is a degree 1 function with 4 parameters, and
$$g_i=\beta_1a_i+\beta_2b_i+\beta_3c_i+\beta_4d_i+\beta_5$$
where $\beta_1,\beta_2,\cdots,\beta_5$ are coefficients, $1\leq i\leq 5$.
\begin{equation*}
\begin{split}
0.4198&=\beta_1+0.25\beta_2+\beta_3+0.06055\beta_4+\beta_5\\
-0.1551&=\beta_1+\beta_2+1.75\beta_3+0.003906\beta_4+\beta_5\\
18.82&=\beta_1+100\beta_2+0.75\beta_3+0.03125\beta_4+\beta_5\\
0.4561&=10\beta_1+0.25\beta_2+0.5\beta_3+0.02148\beta_4+\beta_5\\
0.0374&=1000\beta_1+0.25\beta_2+1.75\beta_3+0.001952\beta_4+\beta_5
\end{split}
\end{equation*}
By calculation
\begin{align*}
\begin{split}
g&=0.00034a+0.18b-0.52c+5.76d+0.54.
\end{split}
 \end{align*}
Then
$$X_t^\alpha=0.00034\lambda+0.18\mu-0.52\alpha+5.76t+0.54.$$
If we take the average value of t, we have
$$\bar{t}=0.0238276$$
and average value of $X^\alpha_t$, we have
$$\overline{X^\alpha_t}=3.91564.$$
Therefore
$$0.00034\lambda+0.18\mu-0.52\alpha=3.24.$$
We summarise our deviation as
\begin{prop}
The critical link of parameters for self-similarity of the trajectories of $\alpha$-stable Ornstein-Uhlenbeck process is given by the following liner equation
$$0.00034\lambda+0.18\mu-0.52\alpha=3.24.$$
\end{prop}

\subsection{$\alpha$-stable geometric L$\acute e$vy motion}
Similarly, for the triple $(\lambda, \mu,\alpha)$, we are working on determining a critical link of the three parameters $\lambda$, $\mu$ and $\alpha$ towards the similarity of trajectories. The data and calculations have been processed to obtain the degree 1 linear relationship  are as follows.
\begin{center}
 \begin{tabular}{||c  c  c  c  c||}
 \hline
 $\lambda$&$\mu$& $\alpha$& t& $X^\alpha_t$ \\ [0.5ex]
 \hline
 1&0.5&1.25& 0.001952&1.043\\
 \hline
 1&1&1&0.007813&1.372\\
 \hline
100&0.5&1.75&0.001953&0.9523\\
 \hline
100&10&1.25&0.005859&0.5114 \\
 \hline
1000&1&0.75&0.001796&-0.7903\\[1ex]
 \hline
\end{tabular}
\end{center}
We have degrees $n=1$, variables $m=4$, so terms= $\Big(
\begin{array}{cc}
1+4 \\
1\\
\end{array}\Big)=5.$ If we have $g=f(a,b,c,d)$ which is a degree 1 function with 4 parameters, and
$$g_i=\beta_1a_i+\beta_2b_i+\beta_3c_i+\beta_4d_i+\beta_5$$
where $\beta_1,\beta_2,\cdots,\beta_5$ are coefficients, $1\leq i\leq 5$. We have
\begin{equation*}
\begin{split}
1.043&=\beta_1+0.5\beta_2+1.25\beta_3+0.001952\beta_4+\beta_5\\
1.372&=\beta_1+\beta_2+\beta_3+0.007813\beta_4+\beta_5\\
0.9523&=100\beta_1+0.5\beta_2+1.75\beta_3+0.001953\beta_4+\beta_5\\
0.5114&=100\beta_1+10\beta_2+1.25\beta_3+0.005859\beta_4+\beta_5\\
-0.7903&=1000\beta_1+\beta_2+0.75\beta_3+0.001796\beta_4+\beta_5
\end{split}
\end{equation*}
By calculation
\begin{align*}
\begin{split}
g&=-0.0017124a-0.066287b+0.15752c+68.508d+0.74723.
\end{split}
 \end{align*}
Then
$$X_t^\alpha=-0.0017124\lambda-0.066287\mu+0.15752\alpha+68.508t+0.74723.$$
If we take the average value of t, we have
$$\bar{t}=0.0038746$$
and average value of $X^\alpha_t$, we have
$$\overline{X^\alpha_t}=0.61768.$$
Therefore
$$-0.0017124\lambda-0.066287\mu+0.15752\alpha=-0.3949911.$$

\begin{prop}
The critical link of parameters for self-similarity of the trajectories of $\alpha$-stable geometric L$\acute e$vy motion is given by the following liner equation
$$-0.0017124\lambda-0.066287\mu+0.15752\alpha=-0.3949911.$$
\end{prop}

\begin{remark}
Here we only consider linear Lagrange interpolation. One can extend to higher order polynomial interpolation in which more computation is needed. Our consideration gives a simple yet efficient calculation.
\end{remark}

\begin{appendices}

\section{$\alpha$-stable random variable generator}\label{code}
Following codes are used to generate sample trajectories \cite{vm}.
\begin{lstlisting}
function r = stblrnd(alpha,beta,gamma,delta,varargin)

if nargin < 4
    error('stats:stblrnd:TooFewInputs','Requires at least four
    input arguments.');
end
if alpha <= 0 || alpha > 2 || ~isscalar(alpha)
    error('stats:stblrnd:BadInputs',' "alpha" must be a scalar
    which lies in the interval (0,2]');
end
if abs(beta) > 1 || ~isscalar(beta)
    error('stats:stblrnd:BadInputs',' "beta" must be a scalar
    which lies in the interval [-1,1]');
end
if gamma < 0 || ~isscalar(gamma)
    error('stats:stblrnd:BadInputs',' "gamma" must be a
    non-negative scalar');
end
if ~isscalar(delta)
    error('stats:stblrnd:BadInputs',' "delta" must be a scalar');
end

[err, sizeOut] = genOutsize(4,alpha,beta,gamma,delta,varargin{:});
if err > 0
    error('stats:stblrnd:InputSizeMismatch','Size information is
    inconsistent.');
end

if alpha == 2
    r = sqrt(2) * randn(sizeOut);

elseif alpha==1 && beta == 0
    r = tan( pi/2 * (2*rand(sizeOut) - 1) );

elseif alpha == .5 && abs(beta) == 1
    r = beta ./ randn(sizeOut).^2;

elseif beta == 0
    V = pi/2 * (2*rand(sizeOut) - 1);
    W = -log(rand(sizeOut));
    r = sin(alpha * V) ./ ( cos(V).^(1/alpha) ) .* ...
        ( cos( V.*(1-alpha) ) ./ W ).^( (1-alpha)/alpha );

elseif alpha ~= 1
    V = pi/2 * (2*rand(sizeOut) - 1);
    W = - log( rand(sizeOut) );
    const = beta * tan(pi*alpha/2);
    B = atan( const );
    S = (1 + const * const).^(1/(2*alpha));
    r = S * sin( alpha*V + B ) ./ ( cos(V) ).^(1/alpha) .* ...
       ( cos( (1-alpha) * V - B ) ./ W ).^((1-alpha)/alpha);

else
    V = pi/2 * (2*rand(sizeOut) - 1);
    W = - log( rand(sizeOut) );
    piover2 = pi/2;
    sclshftV =  piover2 + beta * V ;
    r = 1/piover2 * ( sclshftV .* tan(V) - ...
        beta * log( (piover2 * W .* cos(V) ) ./ sclshftV ) );

end

if alpha ~= 1
   r = gamma * r + delta;
else
   r = gamma * r + (2/pi) * beta * gamma * log(gamma) + delta;
end

end
\end{lstlisting}

\section{Sample trajectories of $\alpha$-stable Ornstein-Uhlenbeck process}\label{trajec1}
\begin{figure}[H]
\centering
\includegraphics[width=1.24\textwidth, angle =90 ]{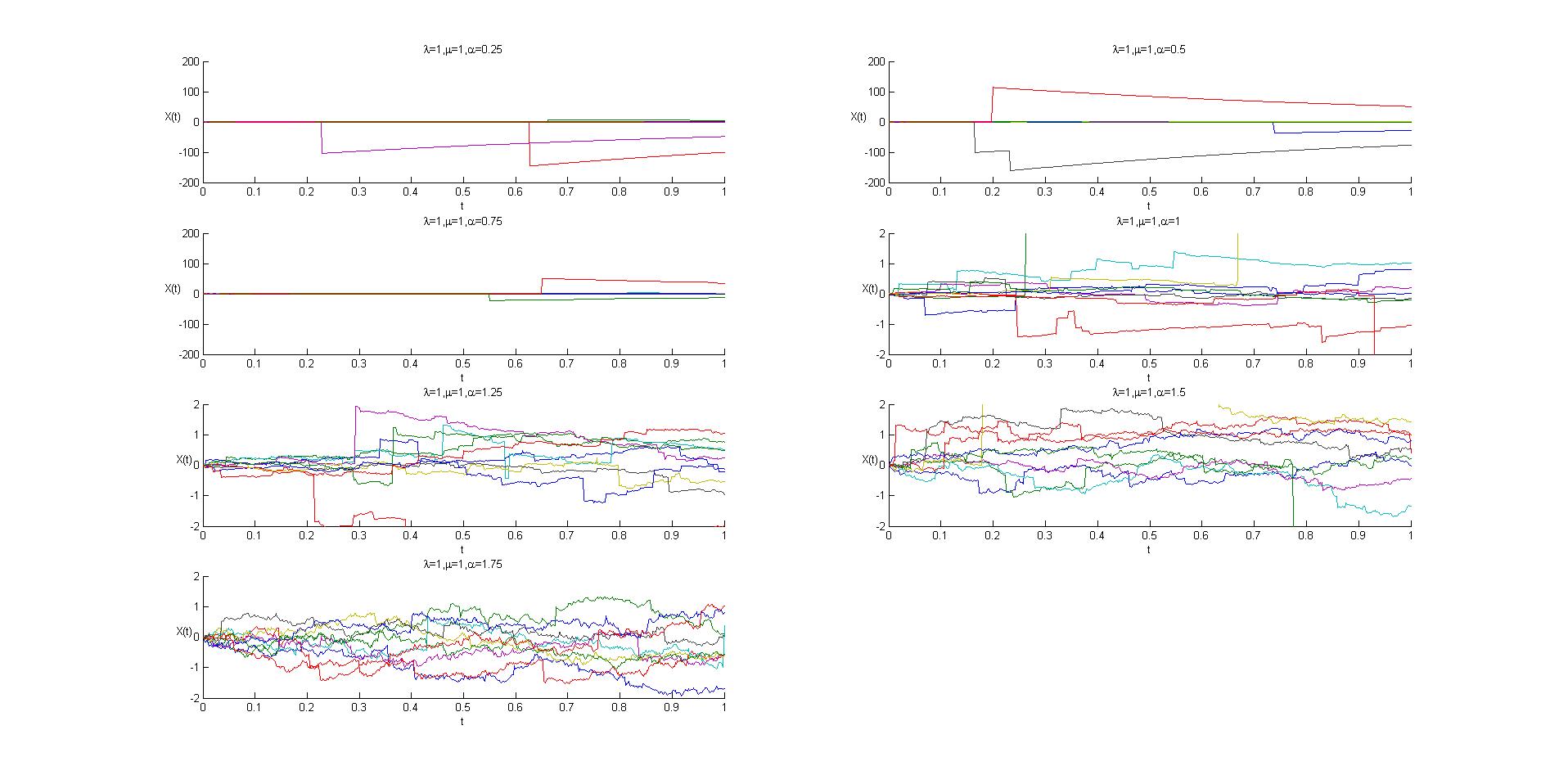}
\caption{Fix $\lambda$=1 and $\mu$=1 with $\alpha$ increases}
  \label{lambda1mu1OU}
\end{figure}

\begin{figure}[H]
\centering
\includegraphics[width=1.4\textwidth, angle =90 ]{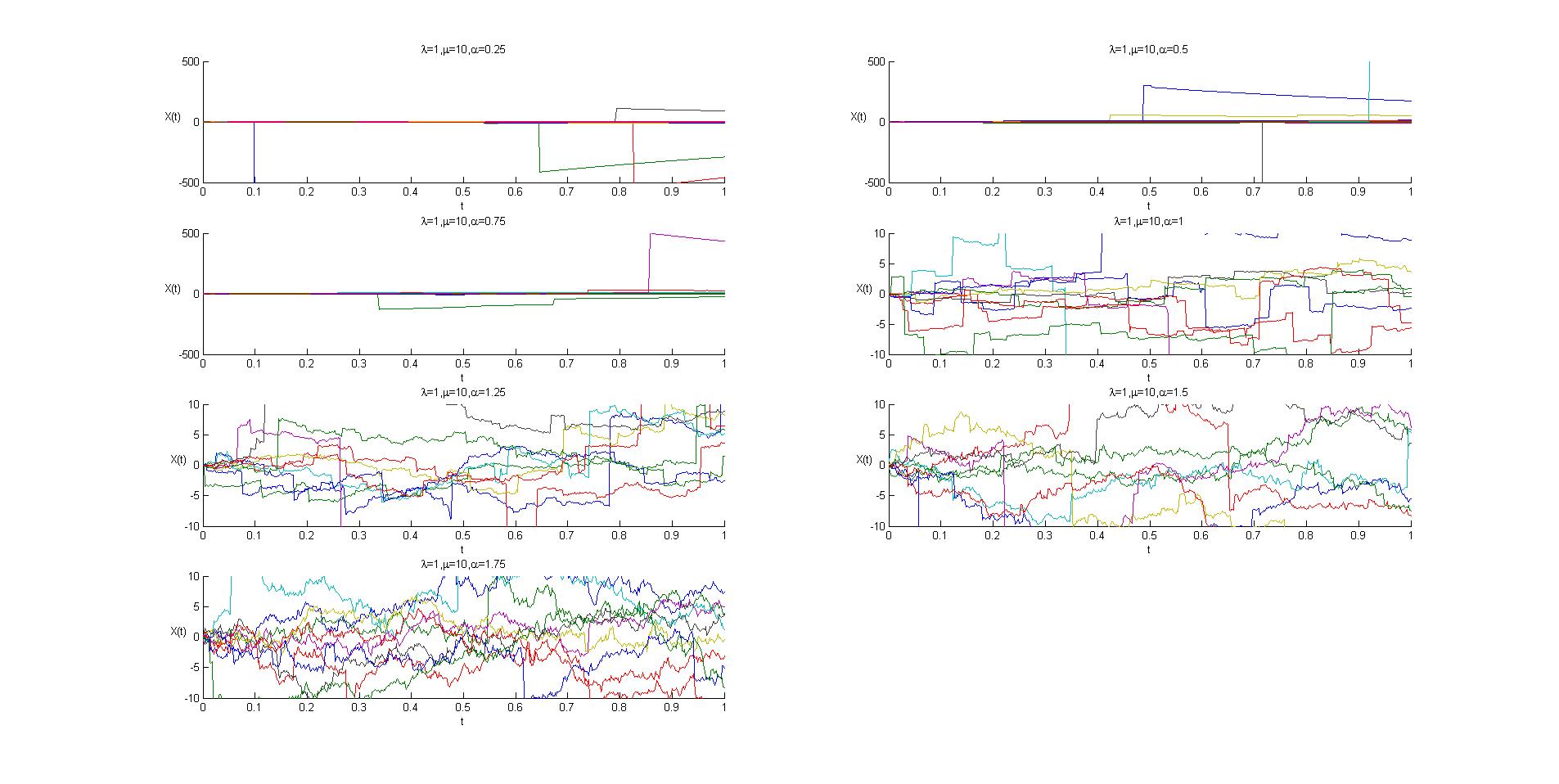}
\caption{Fix $\lambda$=1 and $\mu$=10 with $\alpha$ increases}
  \label{lambda1mu10OU}
\end{figure}

\begin{figure}[H]
\centering
\includegraphics[width=1.4\textwidth, angle =90 ]{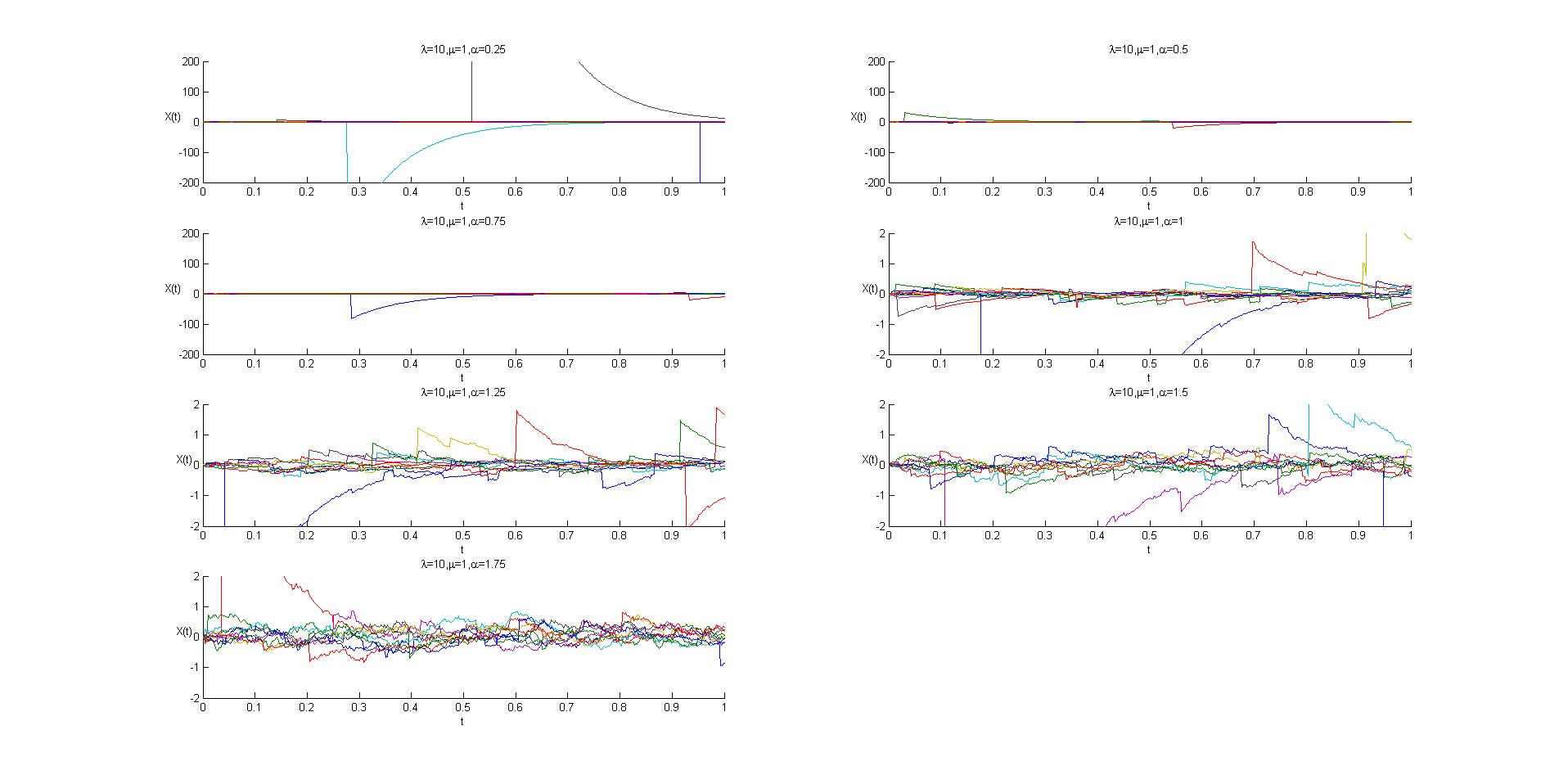}
\caption{Fix $\lambda$=10 and $\mu$=1 with $\alpha$ increases}
  \label{lambda10mu1OU}
\end{figure}

\begin{figure}[H]
\centering
\includegraphics[width=1.4\textwidth, angle =90 ]{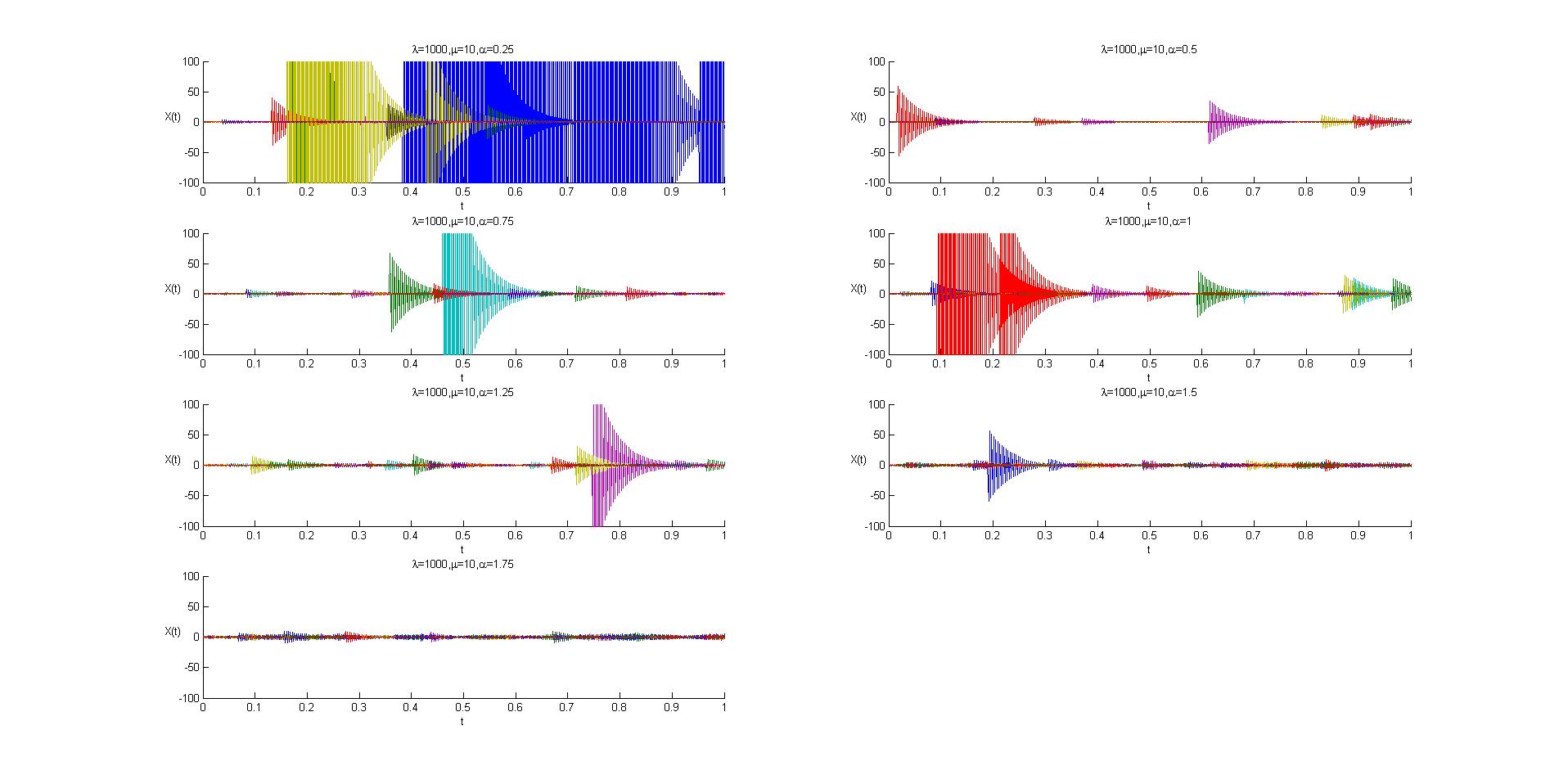}
\caption{Fix $\lambda$=1000 and $\mu$=10 with $\alpha$ increases}
  \label{lambda1000mu10OU}
\end{figure}
\section{Sample trajectories of $\alpha$-stable geometric L$\acute e$vy motion}\label{trajec2}
\begin{figure}[H]
\centering
\includegraphics[width=1.24\textwidth, angle =90 ]{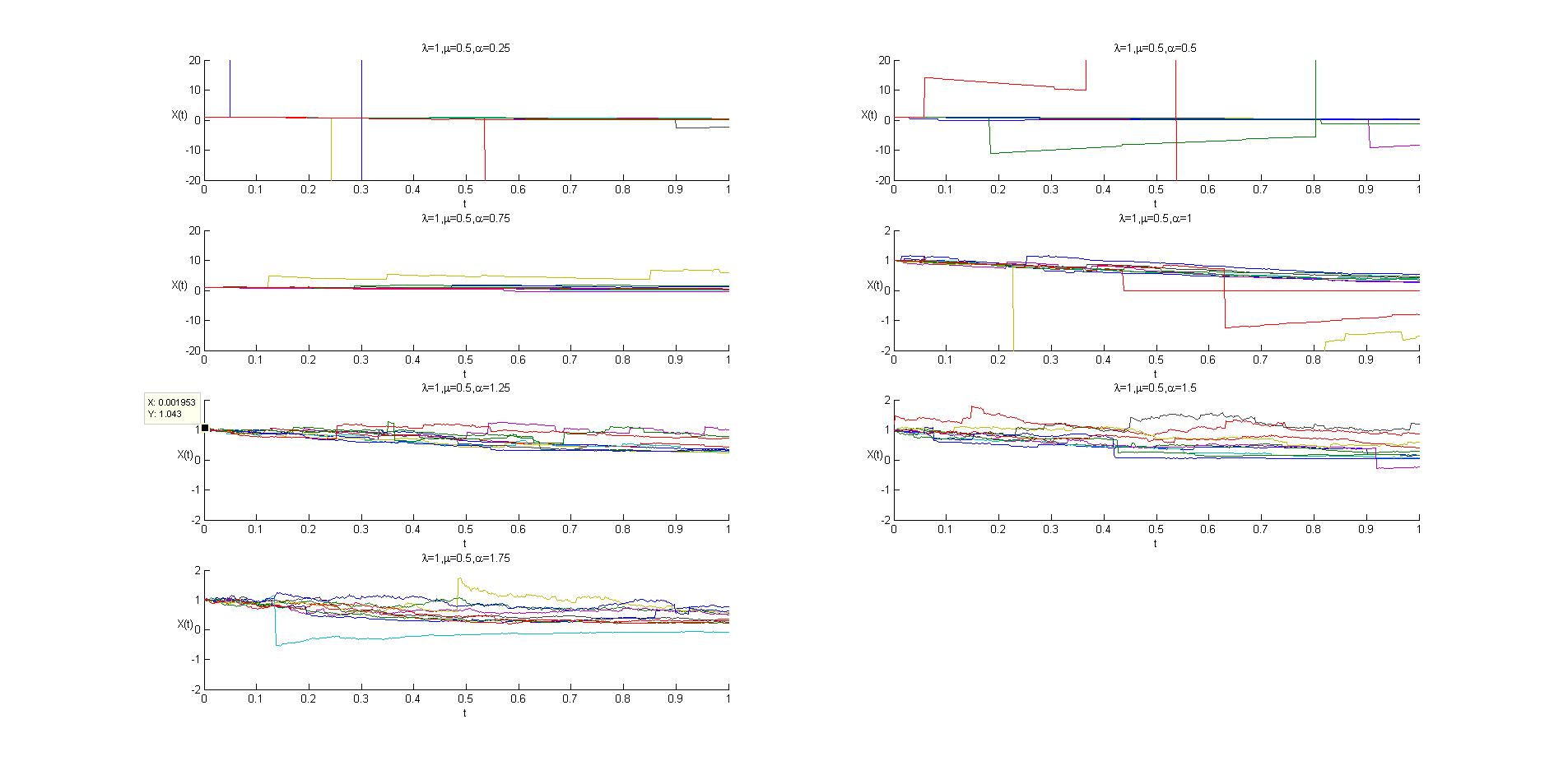}
\caption{Fix $\lambda$=1 and $\mu$=0.5 with $\alpha$ increases}
  \label{lambda1mu05}
\end{figure}

\begin{figure}[H]
\centering
\includegraphics[width=1.4\textwidth, angle =90 ]{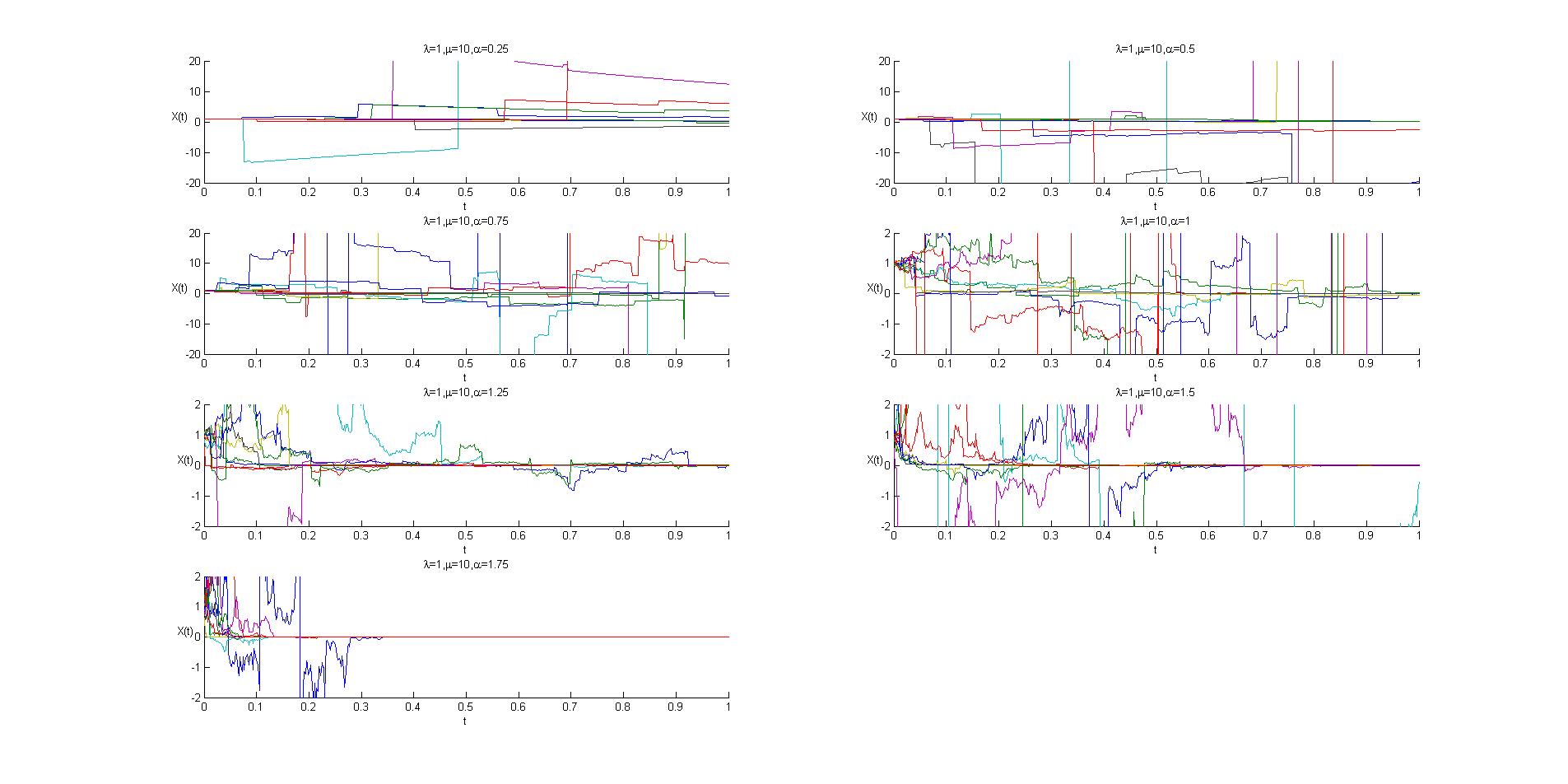}
\caption{Fix $\lambda$=1 and $\mu$=10 with $\alpha$ increases}
  \label{lambda1mu10}
\end{figure}

\begin{figure}[H]
\centering
\includegraphics[width=1.4\textwidth, angle =90 ]{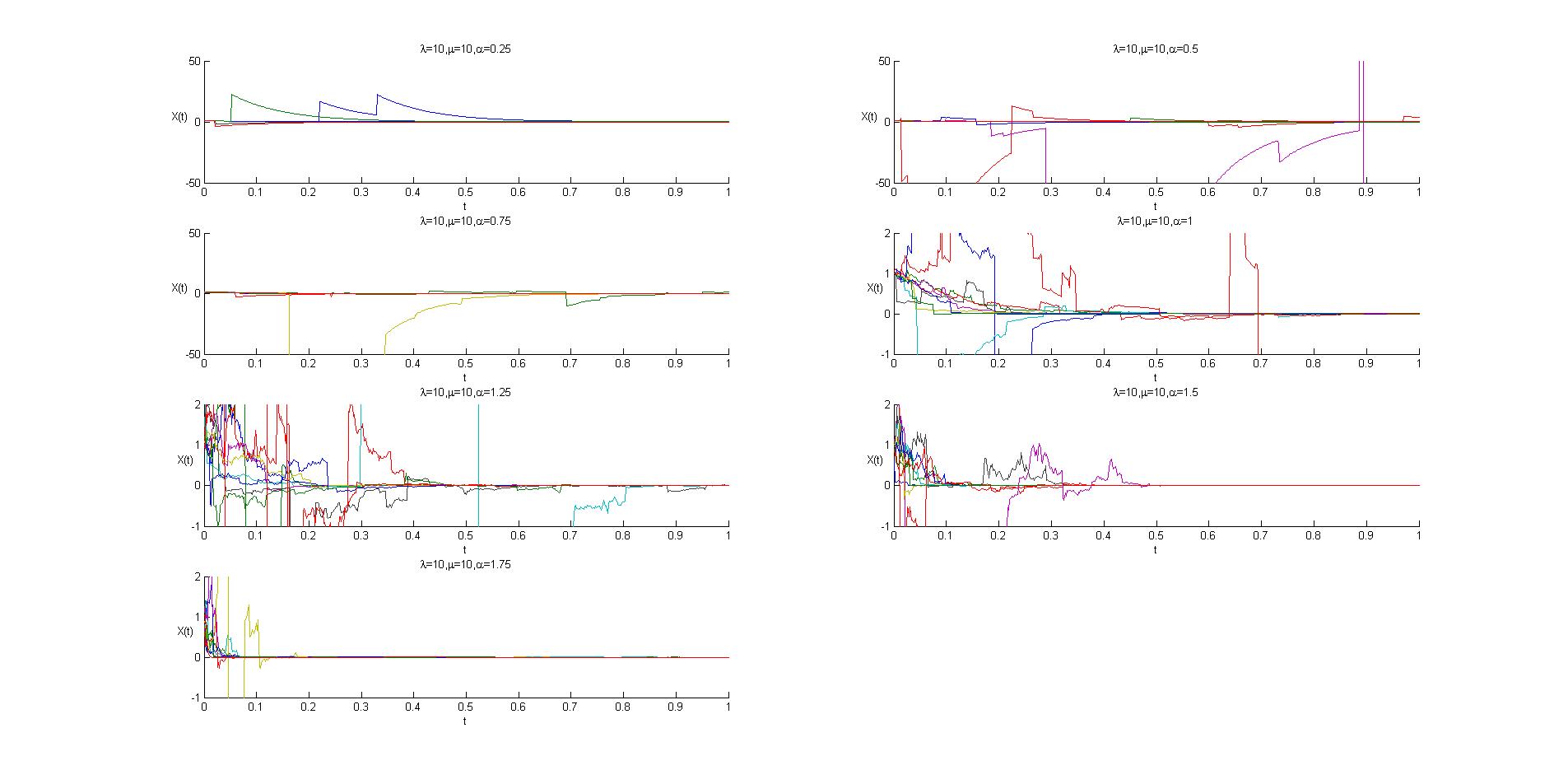}
\caption{Fix $\lambda$=10 and $\mu$=10 with $\alpha$ increases}
  \label{lambda10mu10}
\end{figure}

\begin{figure}[H]
\centering
\includegraphics[width=1.4\textwidth, angle =90 ]{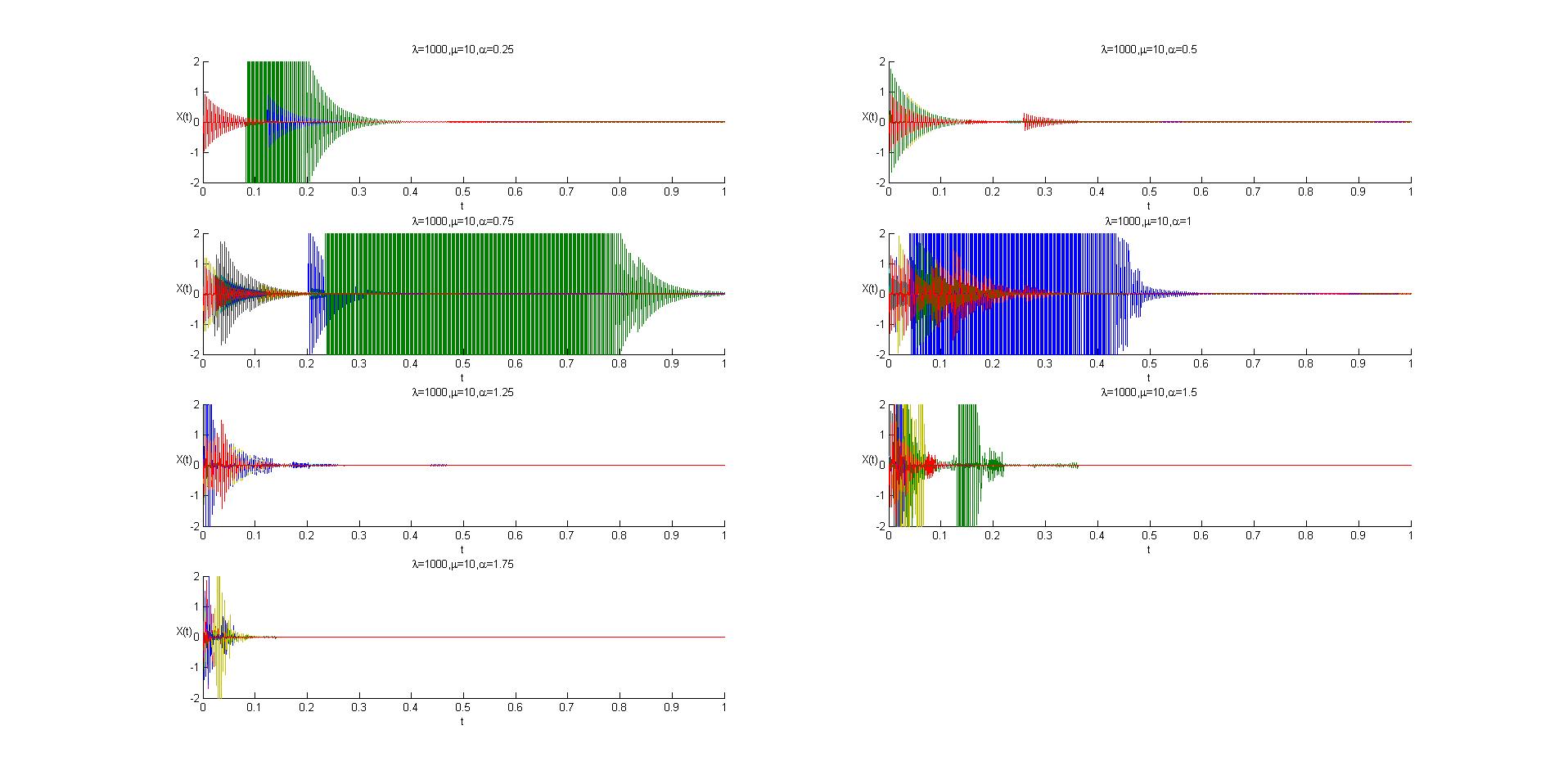}
\caption{Fix $\lambda$=1000 and $\mu$=10 with $\alpha$ increases}
  \label{lambda1000mu10}
\end{figure}

\end{appendices}

\end{document}